\documentstyle[12pt,epsfig,here]{article}
\begin{document}
\newcommand{\beq}{\begin{equation}}
\newcommand{\eeq}{\end{equation}}
\def\beqn{\begin{eqnarray}}
\def\eeqn{\end{eqnarray}}

\newcommand{\Tr}{{\rm Tr}\,}
\newcommand{\E}{{\cal E}}

\newcommand{\ntwo}{${\cal N}=2\;$}
\newcommand{\none}{${\cal N}=1\;$}
\newcommand{\noneh}{${\cal N}=\,
^{\mbox{\small 1}}\!/\mbox{\small 2}\;$}
\newcommand{\vp}{\varphi}
\newcommand{\ve}{\varepsilon}
\newcommand{\pt}{\partial}

\begin{flushright}
FTPI-MINN-07/24, UMN-TH-2613/07 
\end{flushright}

 \vspace{3mm}

\begin{center}
{\bf {\large\bf Some Theoretical Developments in SUSY} \\[1mm]
Talk at the 15th International Conference on Supersymmetry\\[1mm]
and the Unification of Fundamental Interactions --- SUSY 07\\[1mm]
Karlsruhe, Germany
July 26 - August 1, 2007}

 \vspace{5mm}
 
{\large M. Shifman}

 \vspace{3mm}
 
{\it  William I. Fine Theoretical Physics Institute,
University of Minnesota,
Minneapolis, MN 55455}

 \vspace{3cm}
 
 {\em Abstract}

\end{center}

\vspace{2mm}

I review theoretical developments 
of the last year or so in nonperturbative supersymmetry.

 \vspace{3mm}
 
\underline{Topics:}

\vspace{3mm}
 
$\bullet$ Metastable vacua at strong coupling;

$\bullet$ Non-Abelian flux tubes, confined monopoles

$\bullet$ 
One step beyond Seiberg's duality 

$\bullet$ Planar equivalence 

$\bullet$ B theory (multileg/multiloop amplitudes) 

$\bullet$ ${\cal N}= (2,0)$ sigma model (Heterotic flux tubes of Edalati-Tong)

\newpage

\section{Metastable supersymmetry-breaking \\ vacua in SQCD}
\label{msbv}

I will start from a topic which lies half-way between purely theoretical aspects
of supersymmetry (SUSY)  and SUSY-based model-building.
As well known, it is not easy to break supersymmetry dynamically
in such a way that no unwanted phenomenological consequences occur.
Usually one has to deal with contrived schemes
which are not elegant, to put it mildly.
A few mechanisms of dynamical SUSY breaking were discovered in 1980's
and '90s (for a review see \cite{SV}). Approximately at the same time people realized
that some generalized Wess-Zumino models can contain, in addition to supersymmetric vacua,
local minima with positive energy density. If the barrier between the latter
and  supersymmetric vacua is high enough they  represent 
long-lived metastable vacua
in which supersymmetry is spontaneously broken \cite{79}.
The models that were built on such vacua in 1990s tended to be rather
awkward.

In 2006 Intriligator, Seiberg and Shih
showed \cite{ISS}  that metastable dynamical SUSY breaking is much more generic and much simpler than was previously thought. They considered ${\cal N}=1$
SU$(N_c)$ SQCD with $N_c+N$ flavors ($N<N_c/2$) as the starting microscopic theory. As usual, each quark flavor is described by two chiral superfields,
$Q^{k,A}$ and $\tilde{Q}_{Ak}$ where $k$ is the color index
($k=1,2, ... ,N_c$) and $A$ is the flavor index ($A=1,2, ... ,N_c+N$).
All quark flavors were endowed by a common mass term $m$
assumed to be much smaller than the dynamical scale parameter $\Lambda$.
This theory, to be referred to as the electric theory, is strongly coupled and 
its dynamics, beyond some general aspects (such as the number
of SUSY vacua), is not amenable to exhaustive analysis.

The vacuum structure of this model
can be studied through its magnetic Seiberg's dual which is in the 
infrared free regime. The quark mass term is converted into a crucial
term in the superpotential of the magnetic (macroscopic) theory 
\beq
{\mathcal W} = {\mathcal W}_{\rm tree} + {\mathcal W}_{\rm anom}\,,
\label{superp}
\eeq
where
\beq
{\mathcal W}_{\rm tree} = \tilde{h}_{A,k}\,h^{kB}\, M^A_B -\mu^2 M_A^A\,,
\label{superpp}
\eeq
$\tilde{h}_{A,k}$ and $h^{kB}$ denote dual quarks, $k=1,2, ... ,N$,
and $\mu\propto \sqrt{m}$.

The tree-level superpotential ${\mathcal W}_{\rm tree} $
yields metastable vacua. 
The anomalous part of the superpotential, $ {\mathcal W}_{\rm anom}\propto {\rm Tr}
W^2$, is responsible for $N_c$ vacua that restore supersymmetry.
As long as $m\ll\Lambda$
supersymmetric vacua lie far away from the metastable vacua
and are separated by a huge barrier. The lifetime of the
metastable vacua can be made longer than the Universe lifetime.
The same condition $m\ll\Lambda$ allows one to control the incalculable K\"ahler potential corrections.

When all quark mass terms are equal,
the electric theory possesses a global vector-like
${\rm SU}(N_c+N)\times{\rm U}(1)_B$ symmetry. This is a flavor symmetry which is spontaneously broken down to S$({\rm U}(N)\times{\rm U}(N_c))$
in the metastable vacua. Thus, the latter form a compact moduli space of 
metastable vacua
\beq
{\mathcal V} =\frac{{\rm U}(N_c+N)}{{\rm S}({\rm U}(N)\times{\rm U}(N_c))}\,,
\eeq
to be compared with $N_c$ isolated stable vacua.

The Intriligator-Seiberg-Shih (ISS) finding opens for investigation
a large class of (hopefully simple) phenomenologically relevant models with dynamical SUSY breaking. Contentious issues such as large flavor symmetries
and the absence of an $R$ symmetry, a usual ``bottle neck"
in model-building,  may find  relatively easy solutions \cite{ISS2}.
An aspect of model-building where the ISS finding may prove promising is constructing a``direct mediation" model, i.e. a model where the messengers are an integral part of the SUSY-breaking sector. Since SQCD and its cousins naturally come with large global flavor symmetries, one could imagine gauging such a symmetry and identifying it with the 
Standard Model. Such a model might have different phenomenology compared with ordinary gauge mediation.

The aspect which is of most importance to me in this talk is purely theoretical.
The ISS work raises the question of where small deformations of Seiberg's duality
can lead us. We will see later (Sect. \ref{damm}) that they will lead us pretty far --- to non-Abelian strings and confined monopoles.

\section{Non-Abelian flux tubes and confined monopoles}

Seiberg and Witten \cite{SW} presented the first ever demonstration of the dual Meissner effect
in non-Abelian theory, a celebrated analytic proof of linear
confinement, which caused much excitement in the community.

It took people three years to realize \cite{HSZ}
that the flux tubes in the 
Seiberg--Witten solution are not those we would like to have in QCD.
Hanany, Strassler and Zaffaroni who analyzed in 1997 the chromoelectric
flux tubes in the Seiberg--Witten solution 
showed that these flux tubes are essentially Abelian (of the Abrikosov--Nielsen--Olesen 
type) so that the hadrons they would create would have nothing to do with
those in QCD. The hadronic spectrum would be significantly richer.
And, say, in the SU(3) case, three flux tubes 
in the Seiberg--Witten solution would not annihilate into nothing, as they 
should in QCD ...

Ever since
searches for genuinely non-Abelian
flux tubes and non-Abelian mono\-poles continued,
with a decisive breakthrough in 2003-04. By that time the program of
finding field-theory analogs of all basic constructions
of string/D-brane theory was in full swing.\footnote{This program started from the
discovery of the BPS domain walls in ${\mathcal N}=1$ supersymmetric gluodynamics
\cite{DS}.}
BPS domain walls, analogs of D branes, had been identified in
supersymmetric Yang--Mills theory. It had been demonstrated that
such walls support gauge fields localized on them.
 and BPS saturated 
string-wall junctions had been constructed \cite{SY1}.
And yet, non-Abelian flux tubes, the basic element of the
non-Abelian Meissner effect, remained elusive.

\subsection{Non-Abelian flux tubes}

\begin{figure}
\begin{center}
\psfig{figure=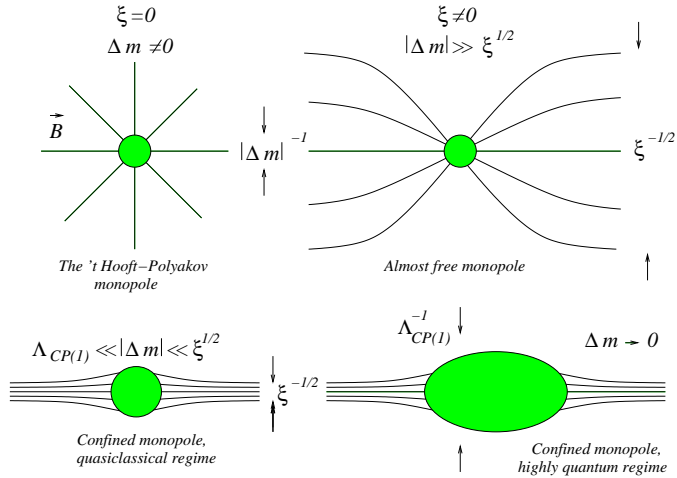,height=2.5in}
\end{center}
\caption{\small Various regimes for monopoles and strings.}
\label{fig:radish}
\end{figure}

They were first found \cite{HT,auzzietal}
in U(2) super-Yang--Mills theories with 
extended supersymmetry, ${\mathcal N}=2$, and two matter hypermultiplets.
If one introduces a non-vanishing Fayet--Iliopoulos parameter $\xi$
the theory develops isolated quark vacua,
in which the gauge symmetry is fully Higgsed, and all elementary excitations are massive.
In the general case, two matter
mass terms allowed by  ${\mathcal N}=2$
are unequal, $m_1\neq m_2$. 
There are free parameters whose interplay
determines dynamics of the theory:
the Fayet--Iliopoulos parameter $\xi$, the mass difference
$\Delta m$ and a dynamical scale parameter
$\Lambda$, an analog of the QCD scale $\Lambda_{\rm QCD}$. 
Extended supersymmetry guarantees that some crucial dependences are holomorphic,
and there is no phase transition.

The number of colors can be arbitrary.
The benchmark model supporting non-Abelian flux tubes
has the gauge group
 SU$(N)\times$U(1) and 
$N$ flavors. The \ntwo vector multiplet
consists of the  U(1)
gauge field $A_{\mu}$ and the SU$(N)$  gauge field $A^a_{\mu}$,
(here $a=1,..., N^2-1$), and their Weyl fermion superpartners
($\lambda^{1}$, $\lambda^{2}$) and
($\lambda^{1a}$, $\lambda^{2a}$), plus
complex scalar fields $a$, and $a^a$. The latter are in the adjoint
representation of SU$(N)$.  In this sector the  global SU(2)$_R$ symmetry inherent to
 \ntwo   models manifests itself through rotations
$\lambda^1 \leftrightarrow \lambda^2$.

The quark multiplets of  the SU$(N)\times$U(1) theory consist
of   the complex scalar fields
$q^{kA}$ and $\tilde{q}_{Ak}$ (squarks) and
the  Weyl fermions $\psi^{kA}$ and
$\tilde{\psi}_{Ak}$, all in the fundamental representation of 
the SU$(N)$ gauge group
($k=1,..., N$ is the color index
while $A$ is the flavor index, $A=1,..., N$).
The scalars $q^{kA}$ and ${\bar{\tilde q}}^{\, kA}$
form a doublet under the action of the   global
SU(2)$_R$ group. 
Quarks and squarks have a U(1) charge too.
A U(1) Fayet-Iliopoulos (FI) term is introduced, with the FI parameter $\xi$.

Both gauge and flavor symmetries of the model are
 broken by the condensation of scalar
fields. A global diagonal
combination of color and flavor groups, SU$(N)_{C+F}$, survives the breaking
(the subscript $C+F$
means a combination of global color and flavor groups). While SU$(N)_{C+F}$ is
the symmetry of the vacuum, the flux tube solutions break it spontaneously.
This gives rise to orientational moduli. Topological stability of the
non-Abelian strings in this model is due to the fact
that $\pi_1({\rm U}(N)/Z_N) = Z_N\times Z$.

\subsection{Confined monopoles}

As various parameters vary, this theory evolves in a very
graphic way, see Fig.~\ref{fig:radish}. At $\xi=0$ but 
$\Delta m \neq 0$
(and $\Delta m \gg \Lambda$) it presents a very clear-cut example
of a model with the standard 't Hooft--Polyakov monopole.
The monopole is free to fly --- the flux tubes are not yet formed.

Switching on $\xi\neq 0$ traps the magnetic fields inside
the flux tubes, which are weak as long as $\xi\ll\Delta m$.
The flux tubes change the shape of the monopole far away from its core,
leaving the core essentially intact. Orientation of the
chromomagnetic field inside the flux tube is essentially fixed. 
The flux tubes are Abelian.

With $|\Delta m|$ decreasing, 
fluctuations in the orientation of the 
chromomagnetic field inside the flux tubes grow. 
Simultaneously, the monopole 
which no loner resembles the 't Hooft--Polyakov monopole, is 
seen as the string junction.
It acquires a life of its own. 

Finally, in the limit $\Delta m\to 0$
the transformation is complete. A global SU(2) symmetry restores
in the bulk. Orientational moduli
develop on the string worldsheet making it truly non-Abelian.
The string worldsheet theory is CP(1) (CP$(N-1)$ for 
generic values of $N$). 
Two-dimensional CP$(N-1)$ models with four supercharges
are asymptotically free. They have $N$ distinct vacuum states.
The non-Abelian flux tubes have degenerate tensions $T_{\rm st} =2\pi\xi$.
The ANO string tension is $N$ times larger. 

\begin{figure}
\begin{center}
\psfig{figure=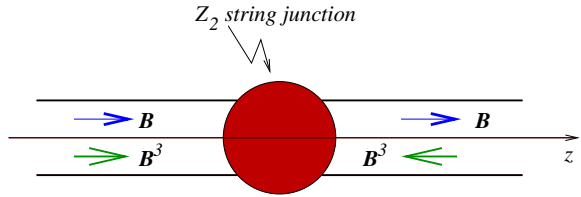,height=1in}
\end{center}
\caption{\small $Z_2$  string junction.}
\label{z2sj}
\end{figure}

Each vacuum state of the worldsheet CP$(N-1)$ theory
presents a distinct string from the standpoint of the bulk theory.
There are $N$ species of such strings. All have the same tension.
Hence, two different strings form a stable junction.
Figure~\ref{z2sj} shows this junction in the limit
\beq
\Lambda_{{\rm CP}(1)}\ll |\Delta m| \ll \sqrt{\xi}
\label{limi}
\eeq
corresponding to the lower left corner of Fig.~\ref{fig:radish}.
The magnetic fluxes of the U(1) and SU(2) gauge groups
are oriented along the $z$ axis. In the limit (\ref{limi})
the SU(2) flux is oriented along the third axis in the internal space.
However, as $|\Delta m| $ decreases, fluctuations
of $B_z^a$  in the internal space grow, and at $\Delta m\to 0$
it has no particular orientation in SU(2) (the lower right corner of Fig.~\ref{fig:radish}).
In the language of the worldsheet theory this phenomenon
is due to restoration of the O(3) symmetry in the quantum vacuum of
the CP(1) model. 

The junctions of degenerate strings present what remains of the
monopoles in this highly quantum regime \cite{SY2,HT2}. It is remarkable that,
despite the fact we are deep inside the highly quantum regime,
holomorphy allows one to exactly calculate the mass of
these monopoles. This mass is given by the expectation value of the
kink central charge in the worldsheet CP$(N-1)$ model
(including the anomaly term). 

What remains to be done?
The most recent investigations zero in on ${\mathcal N}=1$
theories, which are much closer relatives of QCD than ${\mathcal N}=2$.
I have time to say just a few words on the so-called $M$ model suggested recently 
\cite{GSY} which  seems quite promising.
 
 \subsection{{\boldmath$ M$} model}
 
 The unwanted feature of ${\mathcal N}=2$ theory,
 making it less similar to QCD, is the presence of the adjoint scalar field. One can get rid of it making it heavy. To this end we must endow the adjoint superfield
by a mass term. Supersymmetry of the model becomes ${\mathcal N}=1$.
Moreover, to avoid massless modes in the bulk theory (in the limit of very heavy adjoint fields) we must introduce
 a  ``meson"  superfield $M^A_B$
analogous to that emerging in the magnetic Seiberg dual, see Sect.~\ref{msbv},
with an appropriately superpotential.
After the adjoint field is eliminated the theory has  
no 't Hooft--Polyakov monopoles in the quasiclassical limit.
Nevertheless, 
a non-Abelian Meissner effect does take place: condensation of color charges (squarks) gives rise to non-Abelian flux tubes and confined monopoles. The very fact of their existence in ${\mathcal N}=1$ supersymmetric QCD without adjoint scalars was not known previously. The analysis
presented in Ref.~\cite{GSY}  is analytic and is based on the fact that the ${\mathcal N}=1$ theory under consideration can be obtained starting from ${\mathcal N}=2$ SQCD in which the 't Hooft--Polyakov monopoles do exist, through a certain limiting procedure allowing one to track the status of these monopoles at various stages (analogous
to the one described above and summarized in Fig.~\ref{fig:radish}). 

The $M$ model shares many features with the ISS magnetic theory.
I will return to this fact later. Now I would like to note
that non-BPS flux tubes in the
metastable vacua of the SO$( N_c)$ ISS theory
with $N_c+N-4$  flavors were found by Eto et al. \cite{eto1}. 
In a parallel consideration, they made one extra step on the way from
the  SU$( N_c)$ ISS magnetic theory towards the $M$ model. They gauged the baryon
U(1). The U$( N)$ magnetic theory obtained in this way
supports flux tubes in the
metastable vacua \cite{eto1}.

The $M$ model can be regarded as the first cousin of QCD since  the adjoint fields
typical of  ${\mathcal N}=2$
are eliminated in this theory. Even though supersymmetry is considerably weakened,
the overall qualitative picture survives. This is probably
one of the most important findings at the current stage. 

Can a dual of the $M$ model be identified? If yes, 
this would be equivalent to the demonstration of the non-Abelian dual Meissner effect
in ${\mathcal N}=1$.

\section{Dualizing (almost) {\boldmath$ M$} model}
\label{damm}

I started my talk from ISS who slightly mass-deformed
SQCD. This small deformation led to drastic consequences
in the infrared-free magnetic dual theory:
emergence of a non-supersymmetric metastable vacuum.

It turns out that further quite mild deformations of this ``electric" theory
result in a dual ``magnetic" theory which is very close to the $M$ model discussed
above. It preserves all salient features of the $M$ model.

Shifman and Yung considered \cite{SY3}
${\cal N}=1$  
SQCD  with the gauge group U($N_c$) and
$N_c+N$ quark flavors ($N<N_c/2$). 
The U(1) gauge factor gauging baryon charge
is the first (but not last) distinction from ISS.

The next distinction is that we keep $N_c$ flavors 
massless; the corresponding squark fields develop (small) vacuum expectation values
(VEVs) on the Higgs branch. Extra $N$ flavors are endowed with
a mass term $m_q$
which is also small
compared to $\Lambda_Q$, so that all  fields are dynamical
(none can be integrated out). 

\begin{figure}[h]
\epsfxsize=10cm
\hspace{5cm}\epsfbox{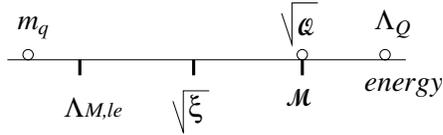}
\caption{\small
Scales of the electric (open points)  and dual magnetic (dashes) theories.}
\label{figscales}
\end{figure}

Within the framework
of this deformation of Seiberg's procedure, on the other side of duality,
the IR free regime is deformed to give rise to
a theory which has the gauge group U$(N$), $N_c$ massive dual-quark flavors
plus $N$ massless dual-quark flavors. In addition to gauge interactions
they are coupled to the meson field $M^A_B$ through a superpotential.
The massive flavors can be integrated out 
while the massless ones develop VEVs. The theory is fully Higgsed.
 The scales relevant to the
electric and magnetic theories are indicated in Fig.~\ref{figscales}.

This set-up leads one from a slightly deformed 
SQCD to the magnetic dual which is very close to the $M$ model: it
supports flux tubes (strings) at weak coupling
and confines non-Abelian (dual) monopoles. The flux tubes are not BPS-saturated,
unlike those of the $M$ model. This is inessential. 

A number of states in the magnetic theory  are light in the sense that
their mass tends to zero in  the limit $m_q\to 0$. Via duality these light states
are in one-to-one correspondence with the light states of the original electric theory.
Thus, duality gets extended to include (in addition to massless moduli)
a part of the spectrum which is light
compared to the natural dynamical scale $\Lambda_Q$ {\em but not massless}.

Extended duality allows one to analyze the magnetic theory at weak coupling
and make a number of highly nontrivial predictions for the quark theory light sector which is at strong coupling.
Non-Abelian monopoles must be important in Seiberg's duality
being related  to ``dual quarks.'' We make one step further
suggesting that the
non-Abelian monopoles of the electric theory {\em are} the ``dual quarks.''
The dual quark fields condense providing (small) masses
to all gauge bosons of the magnetic theory.
The way the magnetic theory is Higgsed is very peculiar ---
it corresponds to baryon-operator dominated vacuum in the  quark theory.
Confined monopoles of the magnetic  theory are to be interpreted as certain
``constituent quarks" of the quark theory. Both form $N$-plets of the
global unbroken SU$(N)$ symmetry which is present in the quark and monopole theories,
on both sides of the extended duality.

In the quark theory color is screened since the theory is fully Higgsed.
There are matter fields in the fundamental representation.
Therefore long strings cannot exist. They are screened/ruptured immediately.
On the dual side we do see strings, however. The scale of the
string-induced confinement $\sqrt{\xi}$ is
small in the original quark theory, much smaller than its dynamical scale,
$\sqrt{\xi} \ll\Lambda_Q$.

This apparent puzzle can be resolved if we assume that
a ``secondary" gauge theory (or a ``gauge cascade") develops
in the original quark theory. Assume that massless composite ``$\rho$ mesons"
whose size is $\sim \Lambda_Q^{-1}$ are formed
in the quark theory which interact with each other via a ``secondary" gauge theory
whose scale parameter is $\sqrt\xi$. At distances $\sim 1/\sqrt\xi$
the above ``$\rho$ mesons" must be viewed as massless gluons.
It is conceivable that they are coupled to
massless ``secondary" quarks which, in addition to their gauge coupling to
``$\rho$ mesons", have nontrivial quantum numbers with respect to
the global SU($N)$. With respect to the original quark theory
the ``secondary" quarks are colorless (``bleached") bound states which include
the original quarks at their core. Their sizes are proportional to $\sim \Lambda_Q^{-1}$
and, hence, they are pointlike on the scale of $\sim 1/\sqrt\xi$,
much in the same way as ``$\rho$ mesons"-gluons.

I think this is a very interesting interpretational issue which calls for further investigation.
On the other hand, some people may want to avoid it. To this end one can follow the road which was suggested recently by Eto et al. \cite{eto2}. Conceptually their strategy
is very similar to that of their predecessors (Ref.~\cite{SY3}), with one 
crucial exception.
Eto et al. base their electric theory on SO$(N_c)$ rather than U$(N_c)$.
Then they endow some quarks (chiral superfields in the vector representation of
SO$(N_c)$) with a mass terms, keeping other flavors massless.
This eliminates metastable ISS vacua. The magnetic theory supports
non-Abelian strings due to the fact that $\pi_1({\rm SO}(N)) =Z_2$. At 
the same time, spinor probe charges
in the electric theory cannot be screened.

\section{Chirally symmetric quasivacuum in \\
supersymmetric gluodynamics?}

Since mid-1980s it is known that
strong- and weak-coupling calculations of the gluino condensate
in supersymmetric gluodynamics do not match.
For SU(2) gauge theory the mismatch is 5/4 (for a review see Ref.~\cite{SV},
Sects. 4.2, 4.3 and 7).

To explain the puzzle Kovner and Shifman suggested \cite{KS}
that an extra chirally symmetric vacuum, with the
vanishing gluino condensate, exists.
This conclusion was also supported by the Veneziano-Yankielowicz effective Lagrangian \cite{VY} with
the simplest kinetic term.
However, later Cachazo et al. proved \cite{cdsw}
that supersymmetric chirally symmetric vacuum is impossible.

The chirally symmetric vacuum may be revived
in a new incarnation, of an unstable (non-supersymmetric) minimum.
This directly follows from ISS. In their set-up
the metastable vacuum is well defined as a manifold
of critical points of ${\mathcal W}_{\rm tree}$. 
In these vacua
the ``dynamically generated part" of the superpotential
\beq
{\mathcal W}_{\rm dyn} \propto \,\left({\rm det}\,M \right)^{1/N} =0\,,
\eeq
cf. Eqs.~(\ref{superp}) and (\ref{superpp}), since $M =0$
in the supersymmetry breaking ISS solution.

The  dynamically generated part
of the superpotential is in fact a low-energy matrix element
of the operator Tr$\lambda^2$, as it follows from the
anomaly relation
\beq
{\mathcal W} = {\mathcal W}_{\rm tree} +\frac{N_c}{16\pi^2}\, {\rm Tr}\, W^2
\eeq
(for a derivation see e.g. Sect. 3.2 of the review \cite{SV}).
Hence, in the metastable ISS vacua
\beq
\langle\,{\rm Tr}\,\lambda\lambda\rangle_{\rm mv} \to 0\,\,\,{\rm  as}\,\,\,
 M\to 0\,,
\eeq
where the subscript mv stands for metastable vacuum.
Making the quark mass $m$ large we approach SUSY gluodynamics.
Apparently, at 
 $m\sim \Lambda $ the metastable vacuum becomes highly unstable.
 Nevertheless,
it may well play the role required from the ``Kovner--Shifman vacuum"
in the strong-coupling instanton calculation of the gluino condensate.

Independent arguments in favor of the chirally symmetric unstable
minimum were given by Douglas, Shelton and Torroba (see Sect. 5 of \cite{DST}). 

\section{Planar Equivalence}
\label{pleq}

Planar equivalence is equivalence in the large-$N$ limit of
distinct QCD-like theories in their common sectors (see \cite{asv}).
Most attention received equivalence between SUSY gluodynamics
and its orientifold  and $Z_2$ orbifold  daughters.
The Lagrangian of the parent theory is
\beq
{\mathcal L}= -\frac{1}{4g_P^2} \, G_{\mu\nu}^a G_{\mu\nu}^a
+\frac{i}{g_P^2}\, \lambda^{a\alpha} D_{\alpha\dot\beta}\bar\lambda^{a\dot\beta}
\eeq
where $\lambda^{a\alpha}$ is the gluino (Weyl) field in the adjoint
representation of SU$(N)$, and $g_P^2$ stands for the coupling constant in the parent theory. The orientifold daughter is obtained by
replacing $\lambda^{a\alpha}$ by the {\em Dirac} spinor in the two-index
(symmetric or antisymmetric) representation (to be referred to as orienti-S
or orienti-AS). The gauge coupling stays intact. 
To obtain the $Z_2$ orbifold daughter (to be referred to as orbi) we must pass
to the gauge group SU$(N/2)\times$SU$(N/2)$, replace $\lambda^{a\alpha}$
by a bifundamental Dirac spinor, and rescale 
the gauge coupling, $g_D^2=2g_P^2$. 

\subsection{Brief history}

 Genesis of planar equivalence can be traced
to string theory. In 1998
Kachru and Silverstein studied  \cite{Kachru:1998ys}
various orbifolds of $R^6$ within the AdS/CFT correspondence, of which I will speak later. Starting from ${\cal N}=4$, they obtained distinct --- but equivalent in the infinite-$N$ limit  ---
four-dimensional daughter gauge field theories with matter, with varying degree of supersymmetry,  all with vanishing $\beta$ functions.\footnote{This statement is
slightly inaccurate; I do not want to dwell on subtleties.}

The next step was made   by Bershadsky et al. \cite{betal}.  These authors
eventually abandoned  AdS/CFT, and string methods at large.
Analyzing gauge field theories {\em per se} they proved
 that  an infinite set of amplitudes 
  in the orbifold daughters of the parent ${\cal N}=4$ theory
 in the large-$N$ limit coincide 
with those of the parent  theory, 
order by order in the gauge coupling. Thus, 
explicitly different theories have the same planar limit, at least perturbatively.

After a few years of relative oblivion, interest in the
issue of planar equivalence was revived by Strassler \cite{stras1}.
He shifted the emphasis away 
from the search   
for supersymmetric daughters, towards engineering QCD-like daughters.
 Strassler considered $Z_N$ orbifolds.
In 2003
an orientifold daughter of SUSY gluodynamics was suggested as
a prime candidate for nonperturbative equivalence \cite{asv1,asv}.
At $N=3$ this orientifold daughter identically reduces to
one-flavor QCD! Thus, one-flavor QCD is planar-equivalent to
SUSY gluodynamics. This remarkable circumstance allows one to copy
results of these theories from one to another.
For instance, color confinement of one-flavor QCD to supersymmetric Yang--Mills,
and the exact gluino condensate in the opposite direction. 
This is how the quark condensate was calculated,
for the first time analytically, in one-flavor QCD \cite{asv2}.

\subsection{Recent Developments}

Kovtun, \"{U}nsal and Yaffe formulated (and derived) \cite{kuy,kuy1}
the necessary and sufficient conditions
for nonperturbative planar equivalence to be valid.  This condition is nonbreaking of 
discrete symmetries: interchange $Z_2$ invariance
for the $Z_2$
orbifold daughter, and $C$ invariance 
for the orientifold daughter.
Although at first glance it does not seem to be a hard
problem to prove that  spontaneous breaking of the discrete symmetries
does not occur, 
in fact, this is a challenging problem which defies exhaustive 
solution so far. 

The question of the discrete symmetry nonbreaking
would be automatically solved if one could
prove that the expansion in fermion loops (say, for the vacuum energy) is
convergent in some sense \cite{asv3}.

To be more exact, let us
 give a mass term $m$ to the fermions, and assume at first this mass
term to be large, $m\gg\Lambda$.  Then the $N_f$ expansion is certainly convergent.
The question is ``is there a singularity, so that at small
$m$ the convergence is lost?"  I believe
that there is no such singularity. If so,
both $Z_2$ orbi and orienti-S/AS
are nonperturbatively equivalent to
supersymmetric gluodynamics. Note that this statement does not refer to $Z_N$ orbi
 with $N>2$.  In this case no mass term is
possible in the orbi theory, it is chiral.

On what I base my belief?
Consider supersymmetric gluodynamics with SUSY slightly broken by
a small mass term of gluino.  At $N=\infty$
the vacuum structure of this theory is exactly the same as the vacuum structure
of pure Yang--Mills (the latter was derived by Witten \cite{Witten:1998uk}, 
see also \cite{afterw,afterw1}).
Thus, I would say that the expansion in the number of fermion loops
should work.
This is of course not a mathematical theorem, but rather a physics argument.

Since for given number of fermion loops and given $m  \neq 0$
each expansion term in supersymmetric gluodynamics
is exactly the same as the corresponding expansion term
in $Z_2$ orbi and orienti-S/AS,
the fermion loop expansions in all three theories must be convergent.

Since in pure gauge theory, with no fermions,  the vacuum is 
unique \cite{Witten:1998uk},
then so is the case for $Z_2$ orbi and orienti-S/AS at $m \neq 0$.
The uniqueness of the vacuum state (for $\theta =0$)
implies the absence of the spontaneous breaking of the discrete symmetries
in the above daughter theories.

If the statement is valid for small $m \neq 0$
extrapolation to $m=0$ must be smooth
since none of these theories has massless particles in the
limit $m=0$. They all have a mass gap 
$\sim\Lambda$.

The studies of planar equivalence reminded us of a remarkable fact:
so far we have no rigorous proof of
the absence of the spontaneous breaking of $C$ invariance
in QCD-like theories \cite{kuy1,asv4}. For $P$ invariance such proof
exists \cite{Vafa:1984xg}, it is iron-clad, but essentially non-dynamical.
Can we find a proof of $C$ invariance? (Of course, in QCD {\em per se}
this fact is well established empirically.)

\subsection{Center-group symmetry and the limit $N\to\infty$}

The planar equivalence between the parent and daughter theories described
in the beginning of Sect.~\ref{pleq} holds not only on $R_4$
but in arbitrary geometry. Therefore, one can compare phase 
diagrams and, in particular, temperature dependences.
This topic was open by Sannino \cite{san}, a thorough discussion
was presented recently by \"{U}nsal \cite{uns}.

There  is a famous Polyakov  criterion regarding
confinement/deconfinement in SU$(N)$ Yang--Mills theories.
If one compactifies $R_4$ into $R_3\times S_1$
and considers the Polyakov line along the compactified direction,
its expectation value may or may not vanish.
If it does not vanish, the $Z_N$ symmetry ---
the center of the gauge SU$(N)$ group ---
is broken. On the other hand, if the Polyakov line vanishes
the $Z_N$ symmetry is unbroken.
The former case corresponds to deconfinement, the latter to confinement.

Introducing quarks may bring in  a problem with this criterion,
since there is no center-group symmetry, generally speaking.
This is in one-to-one correspondence with the fact that
there are no genuine long strings in QCD. They break through 
quark-antiquark pair creation.

Here the $N=\infty$ limit helps. In SU$(N)$ Yang--Mills theory with quarks
in the fundamental representation in the 't Hooft limit
the fundamental quarks decouple, we are left with pure Yang--Mills
which does have center group, and once it is broken, the theory is in the confinement phase.
The Polyakov line is
a good order parameter. The same is valid with regards to supersymmetric gluodynamics
even at finite $N$. Gluinos do not decouple at large $N$,
but they do not ruin the center-group symmetry.

However, consider, for instance, the AS orientifold daughter
of supersymmetric gluodynamics.
At $N=\infty$ two-index antisymmetric fermions
do {\em not}  decouple. There is no apparent center-group 
symmetry in this theory, right?
It is clear that the confinement/deconfinement criterion
as the center-group breaking vs. nonbreaking
is in trouble. Through planar equivalence
we know that at  $N=\infty$ the temperature behavior
of this theory is exactly the same as in supersymmetric gluodynamics
where the Polyakov criterion is perfectly applicable. Where is a way out?

What seems obvious is not always correct.
I want to argue that the center-group symmetry which is not seen at the Lagrangian level in orienti theories,\footnote{For even $N$
there is, of course, an obvious $Z_2$ center group, which is no match
to $Z_N$} in fact, appears {\em dynamically} in the 't Hooft limit.

\begin{figure}
\epsfxsize=4cm
\centerline{\epsfbox{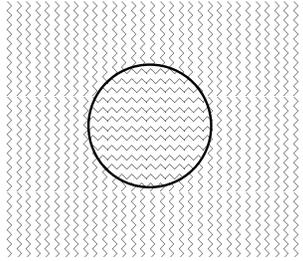}}
\caption{\small
One fermion loop (i.e. $\log \, \det \left( i\not\! \partial + \not\!\!  A^a
\, T^a_{AS} \right)$) in the gluon field background (shown as shaded areas).
}
\label{Mone}
\end{figure}

\begin{figure}
\epsfxsize=6cm
\centerline{\epsfbox{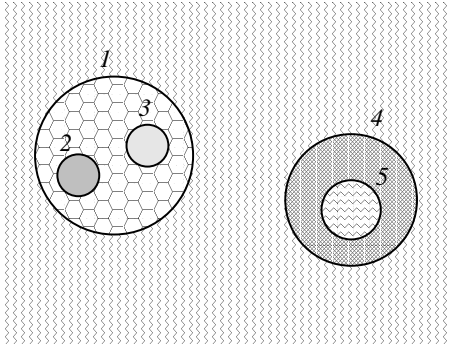}}
\caption{\small An example of fermion multiloops in the gluon field background
at $N\to\infty$.}
\label{Mtwo}
\end{figure}

To see that this is indeed the case let us turn to a ``refined" proof  of planar equivalence
presented in \cite{asv3}. The analysis is based on $N_f$ expansion in the
given ``background" gluon field, with the subsequent integration over the
gluon field. Figure \ref{Mone} displays one-fermion loop in planar geometry (i.e. on a sphere). The gluon fields ``inside" and ``outside"  the loop
do not communicate with each other at $N\to\infty$.
This is indicated by distinct shadings.
Averaging over the gluon field inside the loop is
independent of averaging outside. This means that in calculating this
contribution we can introduce two distinct gluons, two independent SU$(N)$'s.
Each has its $Z_N$ center. However, only one of them survives
due to the fact that the propagating fermion has two fundamental indices upstairs.
A typical multiloop contribution (five fermion loops) is shown in Fig.~\ref{Mtwo}.
Here we have six distinct SU$(N)$ gluons, with five constraints on six $Z_N$
center groups. Again,  one $Z_N$ center survives. Needless to say, this
symmetry disappears at $1/N$ level.


Thus, we do have a $Z_N$ center-group symmetry
in orienti-S/AS! 
Conceptually, this is a non-trivial statement, at the same time being a trivial
consequence of planar equivalence. After all (reduced) SUSY is also not obvious apriori in orienti theories.
This observation invalidates some statements in the literature; in particular,
it restores ``equal-rights" status for even and odd values of $N$.

\section{$D$ Branes in Field Theory}

In 1996 Dvali and I  reanalyzed \cite{DS} supersymmetric gluodynamics,
found an anomalous $(1,0)$
central charge in  superalgebra, not seen at the classical level,
and argued that this central charge will be saturated
by domain walls interpolating
between vacua with distinct values of the order parameter,
the gluino condensate $\langle \lambda\lambda\rangle$,
labeling $N$ distinct  vacua of the theory. 
An exact relation expressing the
wall tension in  terms of the gluino condensate
was obtained. Elementary walls interpolate between 
vacua $n$ and $n+1$, while $k$-walls interpolate
between $n$ and $n+k$.

In  1997 
Witten interpreted \cite{Witten:1997ep} the above BPS walls as analogs of D-branes.
This is because their tension scales as $N\sim 1/g_s$ rather than 
$1/g_s^2$ typical of solitonic objects (here $g_s$ is the string constant).
Many promising consequences ensued. One of them was the Acharya--Vafa
derivation of the wall world-volume theory \cite{AVa}. Using a wrapped $D$-brane
picture and certain dualities they identified the $k$-wall world-volume theory
as 1+2 dimensional U($k$) gauge theory with the field content of
${\cal N}=2$ and the Chern-Simons term at level $N$ breaking ${\cal N}=2$ down to
${\cal N}=1$. Later Armoni and Hollowood exploited this set-up to calculate the 
wall-wall binding energy \cite{AH}.

In 2002  Yung and I considered
${\cal N}=2$ model, weakly coupled in the bulk (and, thus, fully controllable),
which supports both BPS walls and BPS flux tubes \cite{SY1}.
We demonstrated that a gauge field is indeed localized on the
wall; for the minimal wall this is a U(1) field while for nonminimal walls
the localized gauge field is non-Abelian. We also found a BPS wall-string junction
related to the gauge field localization. The field-theory string
does end on the BPS wall, after all!
The end point of the string on the wall, after Polyakov's dualization,
becomes a source of the electric field localized on the wall.
In 2005 Sakai and Tong analyzed \cite{ST} generic wall-string configurations.
Following condensed matter physicists they called them {\em boojums}.

Among advances of the recent years I want to single out
one development: the so-called moduli matrix approach
for description of BPS-saturated multiwall and multistring configurations
and their junctions in ${\mathcal N}=2$ U$(N)$ theories \cite{E1,E2} (for a review see
\cite{E3}). 
The moduli matrix is shown to be a useful tool in exhaustive
description of all BPS solutions since it provides a universal parametrization
for complicated moduli spaces typical in multiwall and multistring configurations. The total moduli space of multiwalls is demonstrated to be given by the complex Grassmann manifold SU$(N_f)/[{\rm SU}(N)\times  {\rm SU}(N_f-N) \times {\rm U}(1)]$.
It can be decomposed into various topological sectors corresponding to boundary conditions specified by particular vacua. Charges characterizing composite solitons give a negative contribution in Abelian theories and can contribute
either positively or negatively) in non-Abelian theories.  Various
applications of the moduli matrix approach were worked out, for instance,
interaction rules of monopoles, vortices, and walls were derived. Among other applications, as a test,
a detailed analysis of the moduli space of  axially symmetric 2-strings  in U(2) gauge theory with two flavors was carried out \cite{Eto:2006cx}.
The moduli space in this case is
${\rm CP}(2)/Z_2$, a manifold containing an $A_1$ -type $(Z_2)$ orbifold singularity.
This limiting case coincides with the answer obtained
in the same problem  previously \cite{Auzzi:2005gr}
by a totally different method.

 
\section {Advanced perturbative calculations
with gluons and quarks}

In gauge theories obtaining  high orders in the perturbative expansion
(multiparton scattering amplitudes at tree level and with loops)
is an immense technical challenge.  Due to the gauge nature
of interactions, the final expressions for such amplitudes
are orders of magnitude simpler than intermediate expressions.

In the 1990's Bern, Dixon and Kosower pioneered applying string methods
to obtain loop amplitudes in  supersymmetric theories. The observed simplicity of these results (generalizing the elegant structure of the Parke-Taylor amplitude \cite{PT})
led to an even more powerful approach based on unitarity.
Their work resulted in an advanced helicity formalism exhibiting a feature 
of the amplitudes, not apparent from the Feynman rules, 
an astonishing simplicity.
In 2003 Witten uncovered \cite{Witten:2003nn} a hidden and elegant mathematical 
structure in terms of algebraic curves in terms of twistor 
variables
in gluon scattering amplitudes: he argued that the 
unexpected simplicity could be understood 
in terms of twistor string theory.

This observation created a diverse and thriving community of theorists advancing perturbative calculations at  tree level and beyond, as it became clear
that loop diagrams in gauge theories have their own hidden symmetry structure.
Most of these results do not directly rely on twistors and twistor string theory,
except for some crucial inspiration. So far, there is no good name for this subject.
Marcus Spradlin noted that an unusually large fraction
of contributors' names
start with the letter B\,.\footnote{E.g. Badger, Bedford, Berger, Bern, Bidder, 
Bjerrum-Bohr,
Brandhuber, Britto, Buchbinder, ... (Of course, one should not forget
about Cachazo, Dixon,  Feng, Forde, Khoze, Kosower, Roiban,  Spradlin, Svr\v{c}ek, Travaglini,  Vaman,  Volovich,  ...). This reminds me of a joke of a proof
given by a physicist that almost all numbers are prime:
one is prime, two is prime, three is prime, five is prime,
while four is an exception just supporting the general rule.
} Therefore, perhaps, we should call it $B$ theory, with B standing for beautiful,
much in the same way as M in $M$ theory stands for magic.
I could mention  a third reason for ``$B$~theory":
 Witten linked the scattering amplitudes to a
topological string known as the ``$B$~model."

$B$ theory revived, at a new level, many methods of the pre-QCD era,
when S-matrix ideas ruled the world. For instance, in a powerful paper due to Britto, Cachazo, Feng and Witten (BCFW) \cite{BCFW}, tree-level on-shell amplitudes were shown in a very simple and general way to obey recursion relations.
Their proof was based only on  Cauchy's theorem and
general (factorization) properties of tree-level scattering!
The BCFW recursion relations gave us a way to calculate
scattering amplitudes without using any gauge fixing or unphysical
intermediate states. 
   
Although the ultimate goal of the $B$ theory is calculating QCD amplitudes,
the concept design of various  ideas and methods is carried out in supersymmetric
theories, which provide an excellent testing ground.
Looking at super-Yang--Mills offers a lot of
insight into how one can  deal with the problems in QCD. 

Of all supersymmetric
theories probably the most remarkable is 
${\cal N}=4$ Yang--Mills. Its special status is due to the fact that (a) it is conformal, and (b) 
in the planar strong coupling limit it is dual to string theory
on AdS$_5\times {\rm S}^5$.

 In 2005 Bern, Dixon and Smirnov calculated
in this theory 
2 gluons $\to$ 2 gluons amplitude up to three loops
\cite{BDS}. Based on this and earlier results
with Anastasiou and Kosower \cite{anak}
they suggested an ansatz for the maximally helicity violating $n$-point amplitudes to all 
loop orders in perturbation theory in the planar limit.
For 2 gluons $\to$ 2 gluons amplitude the Bern-Dixon-Smirnov conjecture takes the form
\beqn
&&{\cal A}(2\,\,{\rm gluons} \to 2\,\,{\rm gluons}) =
{\cal A}(2\,\,{\rm gluons} \to 2\,\,{\rm gluons})_{\rm tree}\times
\nonumber\\[3mm]
&&
\exp \left[ ({\rm IR\,\,\,divergent}) + \frac{f(\lambda )}{8}\left(\ln\frac{s}{t}\right)^2
+{\rm const.} \right]
\label{bds}
\eeqn
where $\lambda$ is the 't Hooft coupling and 
the function $f(\lambda )$ is directly related with the cusp anomalous dimension.

 Recently there was an elegant development in this issue due to Alday and Maldacena
 \cite{AMal}. In a  tour-de-force work 
they performed  the {\em strong} coupling computation by using the 
gauge theory/gravity duality that relates ${\cal N}=4$ Yang--Mills 
to string theory
on AdS$_5\times {\rm S}^5$. They found that the leading order result at large values of the 't Hooft coupling $\lambda$
is given by a single classical string configuration. The classical string solution depends on the momenta $k_i^\mu$ of the final and initial gluons.
The  Alday-Maldacena strong coupling result perfectly matches Eq. (\ref{bds})!
It should be stressed that it is an AdS/CFT match for a rather
non-trivial dynamical quantity.  

\section{${\cal N}= (2,0)$ sigma model (heterotic flux tube)}

Now I want to discuss a stimulating recent development due to Edalati and Tong:
heterotic flux tubes in ${\mathcal N}=1$ theories.
As was mentioned, non-Abelian flux tubes were first discovered in
${\cal N}=2$ SUSY Yang--Mills with the appropriately chosen
matter sector (8 supercharges). 
The flux tube solutions are 1/2 BPS.
Hence, 
the effective low-energy theory of moduli fields
on the string worldsheet must have four supercharges.

In this problem there are two classes of bosonic moduli which split from each other:
orientational moduli and two translational moduli. 
The orientational moduli form CP$(N-1)$ model,
while the translational ones do not interact.
Then the requirement of four supercharges in two dimensions
unambiguously leads us to ${\cal N}= (2,2)$ CP$(N-1)$ model
on the string world sheet. (Of course, in addition,
there is the ${\cal N}= (2,2)$ free theory of translational/supertranslational moduli, which is
dynamically trivial and thus uninteresting.)

An intriguing question arises when one deforms the bulk theory
to break ${\cal N}=2$ down to ${\cal N}=1$.
If this is done in a judicious way, 
e.g. through a superpotential ${\mathcal W}({\mathcal A})$
for the adjoint fields,\footnote{The simplest choice is
${\mathcal W}({\mathcal A})\propto {\mathcal A}^2$. Inclusion of higher powers of
${\mathcal A}$ is possible.}
1/2 BPS flux tube
solutions stay essentially intact.
Moreover, the number of the boson and fermion zero modes,
which become moduli fields on the string worldsheet,
does not change either. However, the bulk theory has now only
four supercharges. According to the standard logic this would imply
two supercharges in the string worldsheet theory.

If superorientational and supertranslational modes
are decoupled, supersymmetrization of the orientational and
translational modes occurs separately.
The orientational modes form  CP$(N-1)$ model.
As well-known, requiring ${\cal N}=1$ SUSY in
CP$(N-1)$ automatically leads to a nonchiral model
with extended supersymmetry, ${\cal N}=(2,2)$. This was the line of reasoning Yung and I followed
in 2005  \cite{Shifman:2005st} in arguing that non-Abelian strings obtained
in ${\cal N}=1$ bulk theories 
with ${\mathcal W}({\mathcal A})\propto {\mathcal A}^2$
have enhanced supersymmetry.

This would be certainly true if the worldsheet theory was just CP$(N-1)$ sigma model.
In fact, it is $C\times {\rm CP}(N-1)$ sigma model. Edalati and Tong 
noted \cite{ET}
that the latter does have a generalization with two supercharges due to the fact
that superorientational and supertranslational modes gets entangled.
In two dimensions there are two distinct superalgebras with two supercharges:
the nonchiral (1,1) algebra and the chiral
${\cal N}= (2,0)$ algebra. It is the latter which was shown \cite{ET} to be relevant.
As a result, the right- and left-moving fermions acquire different interactions, and
the flux tube becomes heterotic!

In essence, the deformation of the ${\rm CP}(N-1)$ model
found by Edalati and Tong reduces to a four-fermion interaction
coupling left-handed superorientational fields $\Psi_1^{\dagger\,\bar j}$ and 
$\Psi_1^{ i}$ with the left-handed fermion fields $\eta_1,\,\,\eta_1^\dagger $
originating from the would-be supertranslational modes
(the two modes corresponding to two
supercharges that are lost in the transition from
${\cal N}=2$ to ${\cal N}=1$). It is of the type
$R_{i\bar j} \, \Psi_1^{ i}\Psi_1^{\dagger\,\bar j}\, \eta_1\eta_1^\dagger  $.
Direct derivation of such terms from the bulk theory is quite tricky.
Instead, Edalati and Tong considered a class of superpotentials 
${\mathcal W}({\mathcal A})$ in the bulk theory, derived the corresponding bosonic sector of the worldsheet theory and then reconstructed the fermionic sector
using  ${\cal N}= (2,0)$. As an independent check
they showed that various symmetries of the bosonic sector of the worldsheet theory
match symmetries of the bulk theory.

Direct derivation of the fermionic sector of string worldsheet theories
for various ${\cal N}=1$ bulk theories supporting
non-Abelian 1/2 BPS flux tubes 
remains an open question.

Very promising developments are expected to follow \cite{Tong:2007qj}.

\section*{Acknowledgments}

I am grateful to Adi Armoni, Zvi Bern, David Shih, David Tong, Mithat \"{U}nsal,
and  Alyosha Yung for valuable discussions. 
I would like to thank Wim de Boer and other organizers of SUSY-07
for kind hospitality in Karlsruhe. 
This work   was
supported in part by DOE grant DE-FG02-94ER408.

\end{document}